\documentstyle[psfig,aps,prl,twocolumn]{revtex}

\begin{document}
\draft \author{Immanuel Bloch, Theodor W.~H\"ansch, and Tilman Esslinger}
\address{Sektion Physik,
Ludwig-Maximilians-Universit\"at, Schellingstr.\ 4/III, D-80799
Munich, Germany and\\ Max-Planck-Institut f\"ur Quantenoptik, D-85748
Garching, Germany}  \title{An Atom Laser with a cw Output Coupler} \maketitle
\begin{abstract}

\noindent
We demonstrate a continuous output coupler for magnetically trapped atoms. Over a period of up to 
\hbox{100 ms} a collimated and monoenergetic beam of atoms is continuously extracted from a Bose-
Einstein condensate. The intensity and kinetic energy of the output beam of this atom laser are 
controlled by a weak rf-field that induces spin flips between trapped and untrapped states. 
Furthermore, the output coupler is used to perform a spectroscopic measurement of the condensate, 
which reveals the spatial distribution of the magnetically trapped condensate and allows 
manipulation of the condensate on a micrometer scale.
\end{abstract}

\pacs{03.75.Fi, 05.30.Jp, 32.80.Pj, 42.55.-f}

\noindent

Four decades ago the first optical lasers\cite{Schawlow} were demonstrated\cite{Maiman,Javan}, 
marking a scientific breakthrough: ultimate control over frequency, intensity and direction of 
optical waves had been achieved. Since then, lasers have found innumerable applications, both for 
scientific and general use. It may now be possible to control matter waves in a similar way, as 
Bose-Einstein condensation has been attained in a dilute gas of trapped 
atoms\cite{Anderson,Bradley,Davis}. In a Bose-Einstein condensate a macroscopic number of bosonic 
atoms occupy the ground state of the system, which can be described by a single wave function. A 
pulsed output coupler which coherently extracts atoms from a condensate was demonstrated 
recently\cite{Mewes,Andrews}. Due to its properties this source is often referred to as an atom 
laser.

In this letter we report on the successful demonstration of an atom laser with a continuous 
output. The duration of the output is limited only by the number of atoms in the condensate. The 
Bose-Einstein condensate is produced in a novel magnetic trap\cite{Esslinger} which provides an 
extremely stable trapping potential. A weak rf-field is used to extract the atoms from the 
condensate over a period of up to 100 ms, thereby forming an atomic beam of unprecedented 
brightness. The output coupling mechanism can be visualized as a small, spatially localized leak 
in the trapping potential. The condensate wave function passes through the leak and forms a 
collimated atomic beam. This continuous output coupler allows the condensate wave function to be 
studied and manipulated with high spatial resolution.

Let us consider a simple model for a cw output coupler \cite{Burnett,Naraschewski,Steck} and 
assume that atoms which are in a magnetically trapped state are transferred into an untrapped 
state by a monochromatic resonant rf-field. In the untrapped state the atoms experience the 
repulsive mean-field potential of the condensate, which can be approximated by the parabolic form 
of a Thomas-Fermi distribution. When the atoms leave the trap the corresponding potential energy 
of the atoms is transformed into kinetic energy. The velocity of the atoms leaving the trap can 
be adjusted by the frequency of the rf-field, because it determines the spatial region where the 
atoms are transferred into the untrapped state.

So far only pulsed output couplers have been demonstrated\cite{Mewes,Kasevich,Helmerson}. If a 
short and hence broadband rf pulse is applied, the output pulse has a correspondingly large 
energy spread. Each portion of the condensate that leaves the trap has a kinetic energy 
distribution with a width comparable to the mean-field energy of the condensate. For this case 
the output coupling process is no longer spatially selective and therefore largely insensitive to 
fluctuations in the magnetic field.

Continuous wave output coupling, as done in this work, can only be achieved if the magnetic field 
fluctuations that the trapped atoms experience are minimized. The level of fluctuations in the 
magnetic field has to be much less than the change of the magnetic trapping field over the 
spatial size of the condensate.

In the experiment we use a novel magnetic trap\cite{Esslinger}, the QUIC-trap, which is 
particularly compact and operates at a current of just 25 A. The compactness of the trap allowed 
us to place it inside a \hbox{$\mu$-metal} box, which reduces the magnetic field of the 
environment and its fluctations by a factor of approximately 100. In combination with an 
extremely stable current supply ($\Delta I / I<10^{-4}$) we are able to reduce the residual 
fluctuations in the magnetic field to a level below \hbox{0.1\,mG}. The rf-field for the output 
coupler is produced by a synthesizer (HP-33120A) and is radiated from the same coil as is used 
for evaporative cooling. The coil has 10 windings, a diameter of \hbox{25\,mm} and is mounted 
\hbox{30\,mm} away from the trap center. The magnetic field vector of the rf is oriented in the 
horizontal plane, perpendicular to the magnetic bias field of the trap.

To obtain Bose-Einstein condensates we use the same set-up and experimental procedure as in our 
previous work\cite{Esslinger}. Typically, 10$^9$ rubidium atoms are trapped and cooled in a 
magneto-optical trap. Then the atoms are transferred into the QUIC-trap where they are further 
cooled by rf-induced evaporation. The QUIC-trap is operated with trapping frequencies of 
\hbox{$\omega_\perp$=2$\pi\times$180 Hz} in the radial and \hbox{$\omega_y=2\pi\times$19 Hz} in 
the axial direction. In this configuration the trap has a magnetic field of \hbox{2.5 G} at its 
minimum.

After the creation of the Bose-Einstein condensate the rf-field used for evaporative cooling is 
switched off, and \hbox{50 ms} later the radio frequency of the output coupler is switched on for 
a time of \hbox{15 ms} in a typical experiment. The field of the output coupler is ramped up to 
an amplitude of \hbox{$B_{\mbox{\tiny rf}}=2.6$ mG} within 0.1 ms. Its frequency follows a linear 
ramp from 1.752 MHz to 1.750 MHz, to account for the shrinking size of the condensate as 
discussed below. Over this period atoms are extracted from the condensate and are accelerated by 
gravity. Subsequently, the magnetic trapping field is switched off and \hbox{3.5 ms} later the 
atomic distribution is measured by absorption imaging. The atom laser output is shown in 
\hbox{Fig.\,1}. The beam contains \hbox{$2\times10^5$} atoms and its divergence in the plane of 
observation is below our experimental resolution limit of \hbox{3.5 mrad}. We obtain an output 
beam over a longer period of time when we reduce the magnetic field amplitude $B_{\tiny 
\mbox{rf}}$ of the rf-field. With absorption imaging we are able to directly image the continuous 
output from the atom laser over 40 ms, with \hbox{$B_{\mbox{\tiny rf}}=1.2$ mG}. A more sensitive 
method is to measure the number of atoms that remain in the condensate after a certain period of 
time. This enables us to monitor the output coupling process over up to 100 ms, with 
\hbox{$B_{\mbox{\tiny rf}}=0.2$ mG}. The magnetic field amplitudes have been calibrated with an 
accuracy of \hbox{20\%}.

It is instructive to estimate the brightness of the beam produced by our atom laser. Defining the 
brightness as the integrated flux of atoms per source size divided by the velocity spreads in 
each dimension \hbox{$\Delta v_x\,\Delta v_y\,\Delta v_z$}\cite{definition spectral brightness}, 
we find that the brightness of our beam has to be at least $2\times 10^{24}$ 
atoms\,s$^{2}$\,m$^{-5}$. To obtain this lower limit for the brightness, we estimate that the 
atomic flux is $5\times 10^6$ atoms/s and that the longitudinal velocity spread is given by 
$\Delta v_z = 3$ mm/s. We further assume a velocity spread $\Delta v_x = 5$ mm/s for the strongly 
confining axis, which corresponds to the chemical potential of the condensate. Our measurements 
show that the velocity spread along the weakly confining axis is less than $\Delta v_y = 0.3$ 
mm/s.

 Assuming a Fourier-limited longitudinal velocity width of $\Delta v_z = 0.3$ mm/s and 
diffraction-limited transverse velocity spreads, a brightness of $4 \times 10^{28}$ 
atoms\,s$^{2}$\,m$^{-5}$ can be reached. Both numbers show that continuous output coupling from a 
condensate creates an atomic beam with a brightness that is orders of magnitude higher than that 
of a state-of-the-art Zeeman slower \cite{meschede} with a brightness of $2.9 \times 10^{18}$ 
atoms\,s$^{2}$\,m$^{-5}$ or an atomic source derived from a magneto-optical trap 
\cite{WiemanLVIS} with a brightness of $8.5 \times 10^{16}$ atoms\,s$^{2}$\,m$^{-5}$.

Let us now consider the geometry of the trap with respect to the output coupling mechanism in 
more detail. The magnetic field $B({\bf r})$ gives rise to a harmonic trapping potential which 
confines the
condensate in the shape of a cigar, with its long axis oriented perpendicular to the 
gravitational force. The rf-field of frequency \hbox{$\nu_{\tiny\mbox{rf}}$} induces transitions 
from the magnetically trapped \hbox{$|F=2, m_{{\tiny F}}=2\rangle$} state to the untrapped 
\hbox{$|F=2, m_{{\tiny F}}=0\rangle$} state, via the \hbox{$|F=2, m_{{\tiny F}}=1\rangle$} state. 
Here $F$ denotes the total angular momentum and $m_F$ the magnetic quantum number. The resonance 
condition \hbox{$\frac{1}{2}\,\mu_{\tiny B}|B({\bf r})|=h\nu_{\tiny\mbox{rf}}$}, where 
$\mu_{\tiny B}$ is the Bohr magneton, is satisfied on the surface of an ellipsoid which is 
centered at the minimum in the magnetic trapping field\cite{Breit-Rabi}. Without gravity the 
condensate would have the same center, so that an undirected output could be 
expected\cite{Naraschewski}. The frequency range in which significant output coupling occurs 
would then be determined by the magnetic field minimum \hbox{$B_{\tiny\mbox{off}}$} and by the 
chemical potential $\mu$ of the condensate: \hbox{$\frac{1}{2}\,\mu_{\tiny 
B}B_{\tiny\mbox{off}}\leq h\nu_{\tiny\mbox{rf}}\leq \frac{1}{2}(\mu_{\tiny 
B}B_{\tiny\mbox{off}}+\mu)$}.

Due to gravity, the minimum of the trapping potential is displaced relative to the minimum of the 
magnetic field. With $g$ being the gravitational acceleration, this displacement is given by 
\hbox{$g/\omega_\perp^2$}, which is \hbox{$7.67\,\mu$m} for our trapping parameters. The 
confinement of the trap and hence the spatial size of the condensate remain the same. In this 
geometry, which is illustrated in Fig.\,2, output coupling occurs only at the intersection of the 
displaced condensate with the ellipsoid that is determined by the resonance condition. Atoms 
leaving the condensate therefore experience a directed force which is dominated by gravity in
our experiment and gives rise to a collimated output beam.
The frequency range over which output coupling can be achieved is larger than without gravity, 
because the condensate is shifted into a region of an increasingly stronger magnetic field 
gradient. The frequency interval \hbox{$\Delta\nu =g\sqrt{2\mu m}/(h\omega_\perp)$}, where $m$ is 
the atomic mass, gives the difference in frequency between an rf-field that is resonant with the 
upper edge and an rf-field that is resonant with the lower edge of the condensate, assuming a 
Thomas-Fermi distribution. For our trapping parameters and \hbox{$7\times 10^5\,$rubidium} atoms 
in the condensate this frequency interval is \hbox{$\Delta\nu = 10.2$ kHz}.

We investigate the condensate spectroscopically by measuring the number of atoms which remain in 
the condensate after \hbox{20 ms} of output coupling, for various amplitudes and frequencies of 
the rf-field.  The number of atoms in the condensate is measured by absorption imaging. Before 
the absorption pulse is applied, the condensate is released from the magnetic trap and expands 
ballistically for 10.5 ms. Due to an inhomogeneous magnetic field, which is applied during the 
switch-off period, atoms in the \hbox{$m_F=2$} and \hbox{$m_F=1$} magnetic sublevel are spatially 
separated from each other, allowing us to determine the population in each of the sublevels 
independently (see also Fig.\,1). The experimental parameters are carefully controlled for each 
series of measurements, so that the number of atoms in the condensate does not fluctuate by more 
than 5\%.

In \hbox{Fig. 3} the number of atoms in the condensate is plotted versus the square of the Rabi 
frequency which is induced by the rf-field, for a fixed frequency of \hbox{1.736 MHz}. With 
increasing rf power the population $N$ in the condensate decreases until it approaches a constant 
level $N_0$. This is to be expected, since the size of the condensate shrinks with its 
depopulation until the overlap with the resonance ellipsoid vanishes. Our measurements can 
therefore be explained by the simple rate equation \hbox{$\frac{dN(t)}{dt}=-\Gamma\left[N(t)-N_0 
\right]$}, in which the depopulation rate $\Gamma$ is proportional to the square of the Rabi 
frequency \hbox{$\Omega=g_F\mu_B B/\hbar$}, with $g_F$ being the Land\'e factor. From a fit to 
our experimental data we obtain \hbox{$\Gamma / \Omega ^2=1.2(2)\times10^{-5}$ s}.

Fig.\,4 shows the dependency of the condensate population on the frequency of the rf-field, for 
\hbox{$\Omega=2\pi\times0.7$ kHz}. The measured distribution has a width of \hbox{13.1(5) kHz} 
(10\% values) and stretches over a larger frequency range than the \hbox{10.2 kHz} estimated (see 
above). This is to be expected, since the Thomas-Fermi distribution does not properly take into 
account the decay of the wave function near the outer edge of the condensate. The condensate wave 
function extends beyond the classical radius $R$, determined by the Thomas-Fermi distribution, 
and falls off exponentially on a characteristic length scale, which is given by 
\hbox{$\delta=R/2\left(\hbar \omega_\perp/\mu\right)^{2/3}$}\cite{Dalfovo,Lundh}. For our 
parameters we obtain \hbox{$\delta/R=0.074$}. Our measurements are in good agreement with these 
considerations. The slight asymmetry of the measured curve is caused by the asymmetry in the
number of atoms per resonance ellipsoid.

The results clearly show the high spectral and spatial resolution of our output coupler. From an 
estimated spectral resolution of \hbox{1 kHz} we obtain a spatial resolution of about \hbox{1 
$\mu$m}. A more precise comparison of our measurements with a theoretical 
model\cite{Naraschewski,Steck} should be possible if the magnetic structure of the \hbox{$F=2$} 
spin state is taken into account and if the output coupling process is studied beyond the Thomas-
Fermi approximation.

The method reported in this paper is a starting point to systematically analyze the spectroscopic 
properties of alkali condensates\cite{Killian}. Our experiments also demonstrate that Bose-
Einstein condensates can be manipulated on a micrometer scale with rf-fields. A similar scheme to 
our output coupler could be used to realize a weak link between two condensates which are trapped 
in different internal states\cite{double condensates}. This would provide the means to observe 
Josephson-type oscillations, as discussed in a recent theoretical paper\cite{Williams}.

Studies on the output properties of the atom laser will provide a fascinating field of research. 
It has already been shown that the rf output coupling process preserves first order 
coherence\cite{Andrews}, but more quantitative and detailed measurements are required. One of the 
crucial properties of an optical laser is its second order coherence\cite{Arecchi}. A direct 
measurement of the second order coherence of an atom laser would reveal the intruiging 
consequences of quantum statistics.

There is a wide range of applications for atom lasers in the field of atom optics. It is now 
conceivable to produce diffraction limited atomic beams which could be focused down to a spot 
size of much less than \hbox{1 nm}. Atom lasers will also revolutionize atom interferometers. 
Highly collimated and slow beams of atoms, as demonstrated in this work, make it possible to 
create atom interferometers with large enclosed areas and superior signal to noise ratio, which 
are idealy suited for precision measurements.

\bigskip

\noindent
We would like to thank Jens Schneider for stimulating discussions.

\begin{figure}
\centerline{\psfig{file=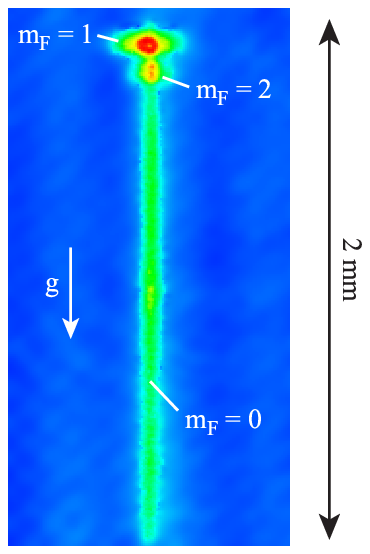,angle=0,width=0.35\textwidth}}
\caption {{Atom laser output: A collimated atomic beam is derived from a Bose-Einstein condensate 
over a \hbox{15 ms} period of continuous output coupling. A fraction of condensed atoms have 
remained in the magnetically trapped \hbox{$|F=2, m_{{\tiny F}}=2\rangle$} and \hbox{$|F=2, 
m_{{\tiny F}}=1\rangle$} state. The magnetic trap has its weakly confining axis in the horizontal 
direction.}}
\label{Fig1}
\end{figure}

\begin{figure}
\centerline{\psfig{file=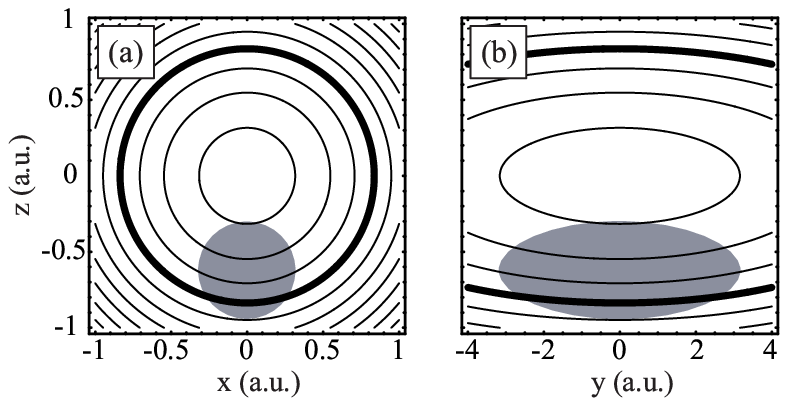,angle=0,width=0.48\textwidth}}
\caption {{Continuous output coupling from a Bose-Einstein condensate. The contour lines 
represent the absolute value of the magnetic trapping field. The thick line indicates the region 
where the rf-field transfers atoms from the magnetically trapped state into an untrapped state. 
Due to gravity, the condensate is trapped \hbox{7.67 $\mu$m} below the minimum in the magnetic 
field.}}
\label{Fig2}
\end{figure}

\begin{figure}
\centerline{\psfig{file=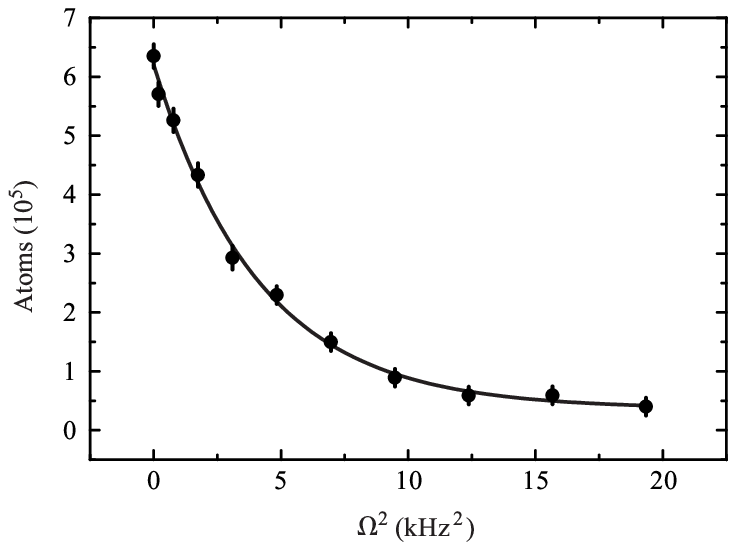,angle=0,width=0.48\textwidth}}
\caption {{Condensate population of the
\hbox{$|F=2, m_{{\tiny F}}=2\rangle$} state after a \hbox{20 ms} period of output coupling, 
versus the square of the Rabi frequency induced by the rf-field. The full line represents a fit 
based on a simple rate equation model (see text).}}
 \label{Fig3}
\end{figure}

\begin{figure}
\centerline{\psfig{file=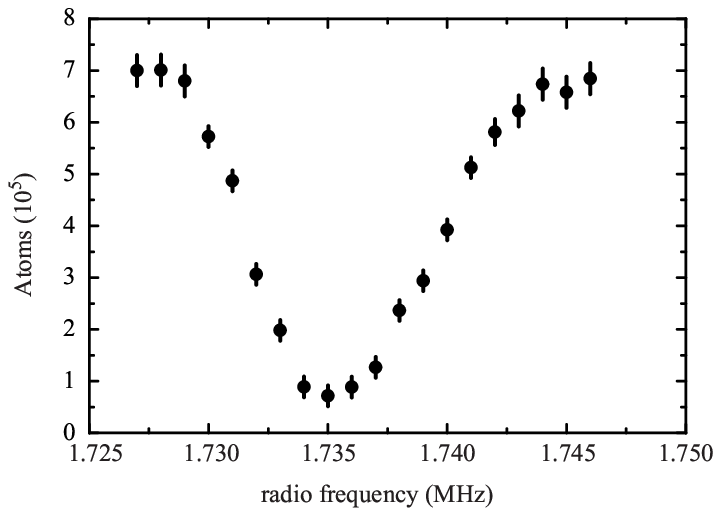,angle=0,width=0.48\textwidth}}
\caption {{Spectroscopy of the Bose-Einstein condensate. The condensate population of the
\hbox{$|F=2, m_{{\tiny F}}=2\rangle$} state is shown for different radio frequencies after a
\hbox{20 ms} period of output coupling. Assuming a
Thomas-Fermi distribution for the condensate density we expect output coupling to
occur within a frequency interval of 10.2 kHz (see text). This does not take into
account the exponential decay of the condensate wave function at the surface of the condensate.}}
\label{Fig4}
\end{figure}


\begin{references}

\bibitem{Schawlow} A. L. Schawlow and C. H. Townes, Phys. Rev. {\bf112}, 1940 (1958).
\bibitem{Maiman} T. H. Maiman, Nature {\bf187}, 493 (1960).
\bibitem{Javan} A. Javan, W. B. Bennett Jr., and D. R. Herriott, Phys. Rev. Lett. {\bf6}, 106 
(1961).
\bibitem{Anderson} M. H. Anderson {\it et al.}, Science {\bf 269}, 198 (1995).
\bibitem{Bradley} C. C. Bradley, C. A. Sackett, J. J. Tollett, and R. G. Hulet, Phys. Rev. Lett. 
{\bf 75}, 1687 (1995) and C. C. Bradley, C. A. Sackett, J. J. Tollett, and R. G. Hulet,
 Phys. Rev. Lett. {\bf 78}, 985 (1997).
\bibitem{Davis} K. B. Davis {\it et al.}, Phys. Rev. Lett. {\bf 75}, 3969 (1995).
\bibitem{Mewes} M. -O. Mewes {\it et al.}, Phys. Rev. Lett. {\bf 78}, 582 (1997).
\bibitem{Andrews} M. R. Andrews {\it et al.}, Science {\bf 275}, 637 (1997).
\bibitem{Esslinger} T. Esslinger, I. Bloch, and T. W. H\"ansch, Phys. Rev. A {\bf 58}, R2664 
(1998).
\bibitem{Burnett}R. J. Ballagh, K. Burnett, and T. F. Scott, Phys. Rev. Lett. {\bf 78}, 1607 
(1997).
\bibitem{Naraschewski} M. Naraschewski, A. Schenzle, and H. Wallis, Phys. Rev. A {\bf 56}, 603 
(1997).
\bibitem{Steck} H. Steck, M. Naraschewski and H. Wallis, Phys. Rev. Lett. {\bf80}, 1 (1998).
\bibitem{Kasevich} B. P. Anderson and M. A. Kasevich, Science {\bf 282}, 1686 (1998).
\bibitem{Helmerson} K. Helmerson {\it et al.} (private communication).
\bibitem{definition spectral brightness}
In this we follow the definition of \cite{Riis} and additionally include the source size of the 
beam, as is usual for the definition of the brightness of a light source in optics \cite{Mandel 
Wolf}.
\bibitem{Riis}E. Riis, D. S. Weiss, K. A. Moler, and S. Chu, Phys. Rev. Lett. {\bf 64}, 1658 
(1990).
\bibitem{Mandel Wolf}L. Mandel and E. Wolf, {\it Optical Coherence and Quantum Optics} (Cambridge 
Univ. Press, Cambridge 1995)
\bibitem{meschede}D. Meschede (private communication).
\bibitem{WiemanLVIS}Z. T. Lu {\it et al.}, Phys. Rev. Lett. {\bf 77}, 3331 (1996).
\bibitem{Breit-Rabi} Using the Breit-Rabi formula we equate that the magnetic splitting between 
the \hbox{$m_F=0$} and \hbox{$m_F=1$} state is 900 Hz larger than the splitting between the 
\hbox{$m_F=1$} and \hbox{$m_F=2$} state.
\bibitem{Dalfovo} F. Dalfovo, L. Pitaevskii, and S. Stringari, Phys. Rev. A {\bf54}, 4213 (1996).
\bibitem{Lundh} E. Lundh, C. J. Pethick, and H. Smith, Phys. Rev. A {\bf55}, 2126 (1997).
\bibitem{Killian} Two-photon optical spectroscopy was recently used to observe BEC in atomic 
hydrogen. T. C. Killian {\it et al.}, Phys. Rev. Lett. {\bf 81}, 3807 (1998) and D. G. Fried {\it 
et al.}, Phys. Rev. Lett. {\bf 81}, 3811 (1998).
\bibitem{double condensates} D. S. Hall {\it et al.}, Phys. Rev. Lett. {\bf 81}, 1539 (1998), D. 
S. Hall, M. R. Matthews, C. E. Wieman, and E. A. Cornell, Phys. Rev. Lett. {\bf 81}, 1543 (1998) 
and M. R. Matthews {\it et al.}, Phys. Rev. Lett {\bf 81}, 243 (1998).
\bibitem{Williams} J. Williams, R. Walser, J. Cooper, E. Cornell, and M. Holland, preprint cond-
mat/9806337.
\bibitem{Arecchi}F. T. Arecchi, E. Gatti, and A. Sona, Phys. Lett {\bf 20}, 27 (1966).

\end{references}
\end{document}